\documentclass{ieeeaccess}
\usepackage{cite}
\usepackage{amsmath,amssymb,amsfonts}
\usepackage{algorithmic}
\usepackage{graphicx}
\usepackage{textcomp}

\newcommand{\bhline}[1]{\noalign{\hrule height #1}}
\usepackage{textgreek}

\usepackage{bm}
\makeatletter
\AtBeginDocument{\DeclareMathVersion{bold}
\SetSymbolFont{operators}{bold}{T1}{times}{b}{n}
\SetSymbolFont{NewLetters}{bold}{T1}{times}{b}{it}
\SetMathAlphabet{\mathrm}{bold}{T1}{times}{b}{n}
\SetMathAlphabet{\mathit}{bold}{T1}{times}{b}{it}
\SetMathAlphabet{\mathbf}{bold}{T1}{times}{b}{n}
\SetMathAlphabet{\mathtt}{bold}{OT1}{pcr}{b}{n}
\SetSymbolFont{symbols}{bold}{OMS}{cmsy}{b}{n}
\renewcommand\boldmath{\@nomath\boldmath\mathversion{bold}}}
\makeatother

\def\BibTeX{{\rm B\kern-.05em{\sc i\kern-.025em b}\kern-.08em
    T\kern-.1667em\lower.7ex\hbox{E}\kern-.125emX}}

\begin{document}
\history{Date of publication xxxx 00, 0000, date of current version xxxx 00, 0000.}
\doi{10.1109/ACCESS.2024.0429000}

\title{Cone-Beam CT Image Quality Enhancement Using A Latent Diffusion Model Trained with Simulated CBCT Artifacts}
\author{\uppercase{Naruki Murahashi}\authorrefmark{1},
\uppercase{Mitsuhiro Nakamura}\authorrefmark{1}, and \uppercase{Megumi Nakao}
\authorrefmark{1},
\IEEEmembership{Member, IEEE}}

\address[1]{Graduate School of Medicine, Kyoto University, Kyoto, 606-8507, JAPAN}

\tfootnote{This research was supported by JSPS Grant-in-Aid for Scientific Research (A) 24H00795 and (B) 23K24282.}

\markboth
{Murahashi \headeretal: Cone-Beam CT Image Quality Enhancement Using A Latent Diffusion Model Trained with Simulated CBCT Artifacts}
{Murahashi \headeretal: Cone-Beam CT Image Quality Enhancement Using A Latent Diffusion Model Trained with Simulated CBCT Artifacts}

\corresp{Corresponding author: Megumi Nakao (e-mail: nakao.megumi.6x@kyoto-u.ac.jp).}

\begin{abstract}
Cone-beam computed tomography (CBCT) images are problematic in clinical medicine because of their low contrast and high artifact content compared with conventional CT images. Although there are some studies to improve image quality, in regions subject to organ deformation, the anatomical structure may change after such image quality improvement. In this study, we propose an overcorrection-free CBCT image quality enhancement method based on a conditional latent diffusion model using pseudo-CBCT images. 
Pseudo-CBCT images are created from CT images using a simple method that simulates CBCT artifacts and are spatially consistent with the CT images. By performing self-supervised learning with these spatially consistent paired images, we can improve image quality while maintaining anatomical structures. 
Furthermore, extending the framework of the conditional diffusion model to latent space improves the efficiency of image processing. 
Our model was trained on pelvic CT-pseudo-CBCT paired data and was applied to both pseudo-CBCT and real CBCT data. The experimental results using data of 75 cases show that with our proposed method, the structural changes were less than 1/1000th (in terms of the number of pixels) of those of a conventional method involving learning with real images, and the correlation coefficient between the CT value distributions of the generated and reference images was 0.916, approaching the same level as conventional methods. We also confirmed that the proposed framework achieves faster processing and superior improvement performance compared with the framework of a conditional diffusion model, even under constrained training settings.
\end{abstract}


\titlepgskip=-21pt

\maketitle 

\section{Introduction}
\PARstart{I}{n} recent years, cone-beam computed tomography (CBCT) has been increasingly used in clinical fields including radiotherapy, in addition to conventional CT. CBCT uses a compact imaging system that consists of an X-ray source and an X-ray detector. Unlike CT, for which the imaging locations tend to be limited, CBCT can be attached to the C-arm of a radiation therapy machine, thereby enabling the immediate acquisition of patient anatomy before and during treatment. Taking advantage of this feature, CBCT is used in radiotherapy to determine the irradiation position. Specifically, this irradiation position is determined by aligning the CBCT image taken immediately before treatment with the planning CT image taken before the treatment start date. However, the planning CT and CBCT images are taken at different dates, and there are differences in the position and shape of organs between the images. Therefore, accurate registration that minimizes these differences is necessary to ensure treatment accuracy. This can be problematical because the quality of CBCT images is generally poor, and large differences in the shapes of organs between CT and CBCT images make accurate registration difficult. Any inaccuracies in the registration create the risk of increased radiation dose to adjacent organs.

The poor quality of CBCT images relative to conventional CT stems from the image acquisition method. In conventional CT, the rotation speed is fast and a tomographic image of a subject is acquired by irradiating fan-shaped X-rays while the X-ray source is rotated around the subject. The three-dimensional structure of the subject is acquired by repeating this process multiple times while moving the subject in the cranio-caudal direction. In contrast, CBCT can acquire the three-dimensional structure of the subject in a single rotation because a cone-shaped X-ray beam is irradiated. However, CBCT has a slower rotation speed, and motion artifacts can occur in the image. Additionally, CBCT is susceptible to scattering. Therefore, the contrast of the image is lower than that of a CT image \cite{Endo2001}, and a wide variety of artifacts, such as streaking and beam-hardening, can occur in the image \cite{Schulze2014}. Because of these factors, the pixel values in CBCT images tend to be less accurate than those in CT images. This makes it difficult to determine the electron density, and CBCT images are not suitable for dose calculation in radiotherapy. Considering these features of CBCT images and the problems in their use, it is desirable to improve the image quality of CBCT images to make it possible to directly calculate the dose to body structures at the time of treatment. Such calculations could be expected to lead to improved treatment accuracy.

Recently, considerable research has examined using deep learning approaches for image translation, and various applications have been reported in the medical imaging field \cite{Yi2019, Kazerouni2023}, such as denoising \cite{Wolterink2017, Yang2018, Zhao2023, Gong2024}, super-resolution \cite{You2020, Xu2024}, and cross-modality translation \cite{Nie2018, Emami2018, Gu2023, Lyu2022} of medical images. The frameworks of the Generative Adversarial Network (GAN) \cite{Goodfellow2014}, CycleGAN \cite{Zhu2017} (an extension of GAN), and conditional diffusion model \cite{Saharia2022, Saharia2023} (an extension of the diffusion model \cite{Ho2020}) are widely used in research for image translations. In particular, the diffusion model has attracted attention because of superior stability compared with GAN, and superior image generation performance comparable to or better than GAN \cite{Dhariwal2021}. Its medical applications are increasing exponentially.

For a conditional diffusion model to learn the task of image-to-image translation, it requires paired image data with one-to-one correspondence before and after translation. However, in clinical practice it is difficult to simultaneously acquire images with and without artifacts or images from different modalities, making it virtually impossible to acquire paired data with perfectly matched structures for the same patient. Therefore, as seen in an example of image quality improvement for head and neck CBCT images \cite{Peng2024}, it is common practice to use a pair of images for which important features have been aligned as a training pair. However, this method is not always effective for the trunk because it is a region where structural changes are likely to occur through the effects of patient posture, respiration, organ motion, and other factors. In such regions, the structures in images may not match sufficiently, even when registration is performed, making it difficult to treat the data as an image pair. Because of these issues, very few studies have used conditional diffusion models to improve the quality of CBCT images of the trunk.

Unlike the conditional diffusion model, CycleGAN follows an unsupervised learning framework that does not require a one-to-one correspondence between images. The method requires only two image groups with different image features and does not even require corresponding numbers of data items between the image groups. It is widely used in fields where it is difficult to acquire paired data, and previous reports describe its effectiveness in reducing dental metal artifacts in CT images and artifacts in CBCT images of the head and neck \cite{Nakao2020, Liang2018}. Methods applying the framework of CycleGAN have been applied to regions subject to structural changes. One example is a study that improved the quality of pelvic CBCT images \cite{Hase2021}. However, because this framework uses data without strict correspondence between the two image groups, there are still issues affecting the maintenance of anatomical structure, such as changes in air pockets and detailed organ geometry before and after image quality improvement. 

In this study, we propose an overcorrection-free CBCT image enhancement method based on a conditional latent diffusion model using pseudo-CBCT images.
In our proposed method, artifacts are artificially added to CT images to create pseudo-CBCT images with the image characteristics of real CBCT images, thereby creating spatially perfectly matched paired data.
By using this paired data for model training, it is possible to improve image quality while maintaining anatomical structures—that is, to achieve image translation without overcorrection.
Moreover, the conditional diffusion model is a model that assumes processing in the image space, and it requires a great deal of computation time and resources when dealing with high-resolution images, such as those obtained in the medical field. 
Our proposed method extends the framework to latent space, making it a conditional latent diffusion model, which reduces computation costs and achieves efficient training and image generation even under constrained settings.

We confirmed the effectiveness of our proposed method for improving the quality of CBCT images through experiments on data from 75 patients undergoing radiotherapy for prostate cancer, evaluating the performance of the model on pseudo-CBCT and real CBCT images. 
In the evaluation of the pseudo-CBCT images, changes in training and generation efficiency resulting from model extension were evaluated in terms of improvement in performance and computation time. 
In the evaluation of real CBCT images, we quantitatively evaluated the changes in CT values and image structures following the translation, evaluating whether the image quality could be improved while maintaining the anatomical structure of the CBCT images. At the same time, we compared the performance with the conventional GAN-based image translation method and the conventional method for learning with aligned real image pairs.

The academic contributions of this study are as follows:

\begin{itemize}
    \item demonstration of the advantages of overcorrection-free image translation achieved through self-supervised learning with spatially consistent data;
    \item analysis of the impact of compression level on the performance and computational efficiency of a diffusion-based approach under constrained settings;
    \item investigation of the performance of the proposed method in improving the quality of CBCT images.
\end{itemize}

\section{Related works}
This section describes deep learning–based methods for reducing artifacts in medical images, focusing on reducing metal artifacts and improving the quality of CBCT images.

\subsection{Metal artifact reduction}
In general, supervised learning for artifact reduction requires paired images with and without artifacts. However, in the case of metal artifact reduction, it is not easy to acquire such paired data for the same patient because there are only a limited number of clinical situations in which in-place metal is later removed. One way to address this problem is to use simulated data for training.

For example, Zhang \emph{et al}. \cite{Zhang2018} proposed a method for a Convolutional Neural Network (CNN) \cite{Krizhevsky2012} framework that uses images with simulated metal items and their artifacts. In their method, artifacts are reduced by correcting the sinogram after the CT image is corrected by the CNN. 
Recently, a new framework based on the diffusion model has been studied. Cai \emph{et al}. \cite{Cai2024} proposed a method to reduce metal artifacts using the cold diffusion \cite{Bansal2023} framework. 
In their study, images were degraded using simulated artifact masks rather than noise. 
However, real artifacts are generally more complex and varied than simulated ones, and the difference can influence performance when these methods are applied to real data.

Another approach other than using simulated data is based on unsupervised learning, which does not require corresponding paired data. Examples of such methods include those based on the CycleGAN framework, such as the artifact disentangle network proposed by Liao \emph{et al}. \cite{Liao2020} and the three dimensional GAN proposed by Nakao \emph{et al} \cite{Nakao2020}. 
In these methods, unsupervised learning was shown to be effective. However, if artifacts are severe or occur over a wide area, they may remain in some parts of the image. In addition, the CycleGAN framework, because of its characteristics, involves the risk of overfitting and translation failure.

\subsection{Improvement of CBCT image quality}
For improving the quality of CBCT images with deep learning, both CBCT and CT images are needed. Unlike the situation for metal artifact reduction, it is easy to acquire these two types of images from the same patient. 
However, it is not possible to acquire the images simultaneously. Therefore, in addition to artifacts, there may be differences in anatomical structure between the images due to differences in the acquisition timing. 
If such images are treated as a pair, the model may learn changes that are not necessary to improve image quality. This can result in anatomical structures being translated in such a way as to deform or create new anatomical structures, leading to a loss of the structural information included in the CBCT images at the time of treatment. 
Considering this, previous studies used data that were aligned between images to minimize differences in anatomical structures for both supervised and unsupervised learning frameworks.

For example, Liang \emph{et al}. \cite{Liang2018} developed a CycleGAN-based framework for head and neck CBCT images that used CBCT images and CT images that were rigidly aligned to the CBCT images. They reported that they achieved translated pixel values close to those of the CT images while maintaining the anatomical structure of the CBCT images. Peng \emph{et al}. \cite{Peng2024} proposed a method using a conditional diffusion model, and they also improved CT values while maintaining structure. However, these studies were performed on the head and neck region where anatomical structures do not change substantially over time, and it has not been shown whether the same methods can be applied to other regions.

For regions such as the trunk that are subject to deformation due to patient posture, respiration, and organ motion, there are many differences between CBCT and CT images. Considering this, Chen \emph{et al}. \cite{Chen2024} improved the image quality of lung CBCT using a conditional diffusion model and data transformed with non-rigid registration to reduce differences between images as much as possible. However, in the abdominal and pelvic regions of the trunk, complex and irregular deformations of organs and associated artifacts tended to obscure tissue contours, decreasing the accuracy of the registration. In addition, the presence or absence of gas varies over time. Therefore, even with non-rigid registration, organ shapes may not match sufficiently, making it particularly difficult to acquire paired data with a strict one-to-one correspondence. For this reason, researchers applying methods to abdominal and pelvic regions have mainly used the CycleGAN framework, which does not require corresponding paired data. 

One example of this is the study by Gao \emph{et al} \cite{Gao2023}. They focused on the problem of severe streaking artifacts caused by abdominal organ motion, which degrade the image quality improvement achievable with CycleGAN, referring to the experimental results of Liu \emph{et al} \cite{Liu2020}. 
To solve this problem, they constructed a GAN-based streaking artifact reduction network (SARN) and used it with CycleGAN. 
SARN was trained on simulated data with streaking artifacts, generated by randomly deforming CT images and reconstructing composite sinograms by combining the sinograms from the pre- and post-deformation images.
Their study showed that images with severe streaking artifacts can be effectively improved by reducing streaking artifacts with SARN, and other artifacts with CycleGAN. 
However, this method, which requires two GANs, raises concerns about the increased risk of overfitting and translation failure.

Other examples of direct translations from CBCT images made using CycleGAN include a study on lower abdominal CBCT images by Kida \emph{et al} \cite{Kida2020}. Their study suggests a problem with the properties of CycleGAN, which is that the translation may fail and result in excessive distortion in the image. 
Hase \emph{et al}. \cite{Hase2021} performed a study on the pelvic region using a set of rigidly aligned CT-CBCT images. Although there were differences in artifacts and detailed organ shapes between the images, there were also many common anatomical structuress. Based on this fact, their method incorporated cycle consistency loss into a GAN-based paired image translation model such as pix2pix \cite{Isola2017}. This enabled image quality improvement without serious translation failures.
However, there are issues concerning the maintaining of structures, such as changes in organ shapes and the creation of new organs.

\section{Methods}

\subsection{Pseudo-CBCT image creation procedure}
\label{subsec:make pseudo-CBCT} 
This section describes creation of the pseudo-CBCT images used in the experiments. The primary objective of this study is to demonstrate the advantages of overcorrection-free image translation enabled by self-supervised learning with spatially consistent data. 
In this study, spatially consistent data are obtained by creating pseudo-CBCT images solely from CT images. Ideally, pseudo-CBCT creation should comprehensively simulate the physical processes of imaging, thereby faithfully reproducing artifacts. However, such comprehensive simulations are difficult to implement and verify. Therefore, in this study, we create pseudo-CBCT images with representative CBCT artifacts using a simple method.

\begin{figure}[!t]
    \centering
    \includegraphics[width=\columnwidth]{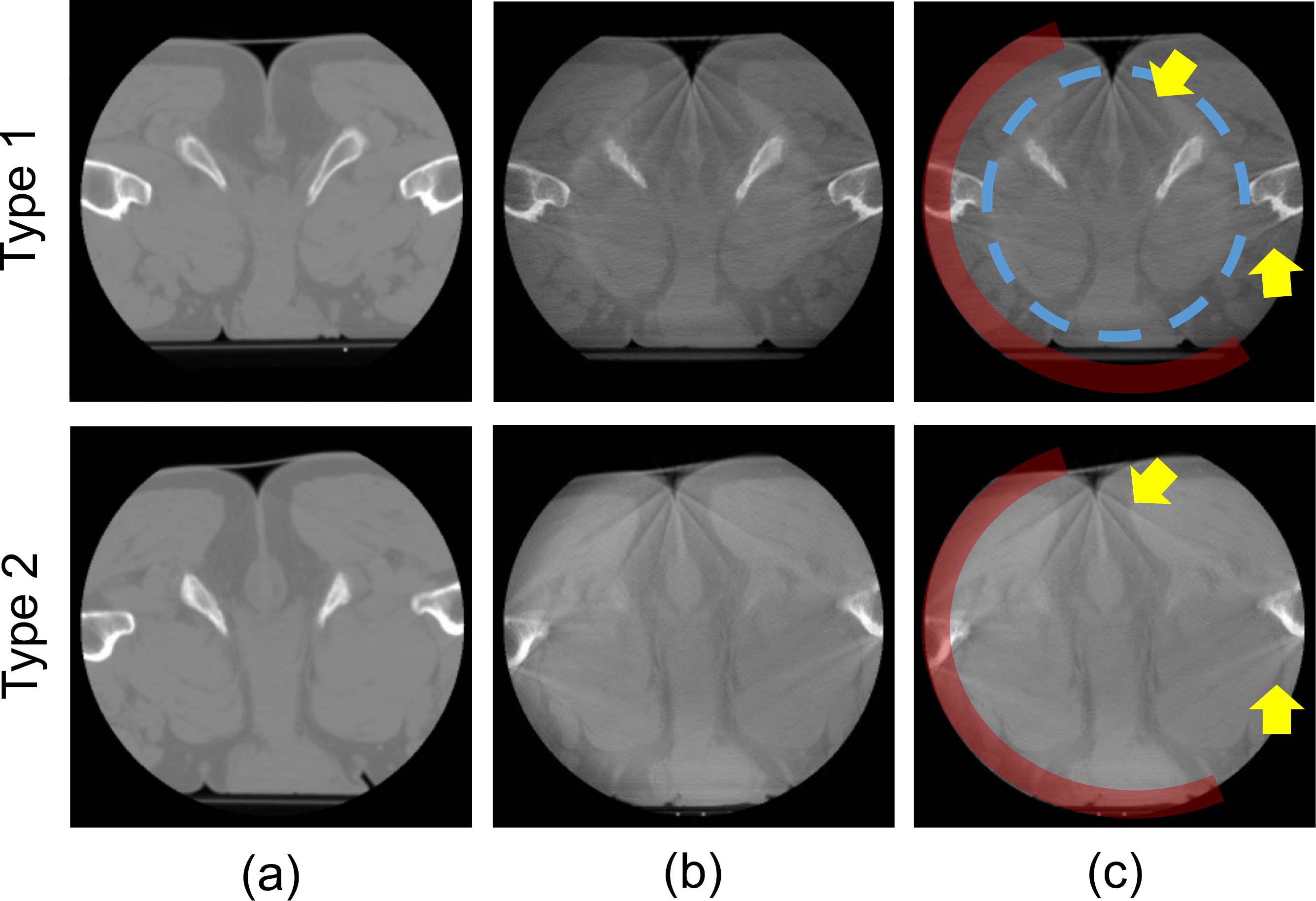}
    \caption{Difference between CT and CBCT images. CBCT images are classified into two main types according to image characteristics: Type 1 is characterized by decreased CT values, while Type 2 is marked by poor contrast. (a) CT image, (b) CBCT image, (c) artifact location on CBCT image. The arrows indicate radial artifacts, and CT values are reduced around the blue dashed line and in the area of the red line.
    }
    \label{fig:CT&CBCT}
\end{figure}

Before explaining the pseudo-CBCT images, we first describe the major features observed in the real CBCT images.
Figure \ref{fig:CT&CBCT} shows two examples of CT images, real CBCT images, and their artifact locations. 
In the real CBCT images used in this study, two major trends of image features can be observed: overall low CT values (Type 1), and poor contrast (Type 2).
As a local feature, a circular region at the center of the image (blue dashed line) is observed only in Type 1. Outside this region, the CT values are further reduced, and there are differences in CT values between the same tissues.
Radial artifacts, as indicated by the arrows, are observed in both Type 1 and Type 2. 
In addition, low CT values at the edge of the imaging field (red line) are also observed in both types, but are more pronounced in Type 2.
These features arise from various factors, among which we focus only on those that tend to produce characteristic artifacts, including scatter, motion artifacts, beam-hardening, and geometric factors.
In this study, the pseudo-CBCT image creation procedure is designed to visually reproduce the observed image characteristics using a simplified process accounting for these factors.

The pseudo-CBCT image creation is modeled as degradation processes in the sinogram and image domains.
Let $s$ denote the sinogram obtained from the CT image.
The degraded sinogram is defined as in equation (\ref{eq:sino_domain}).

\begin{equation}
    \label{eq:sino_domain} 
    \hat{s}=\mathcal{F}(s; \varphi, \sigma, c_0),
\end{equation}

\noindent
where $\varphi$ is a random displacement field, $\sigma$ is the standard deviation of the normal distribution used to create $\varphi$, and $c_0$ is a parameter that determines the degree of contrast adjustment. A larger $\sigma$ results in stronger deformation, and $c_0 = 1.0$ corresponds to no degradation, with larger values of $c_0$ resulting in stronger degradation.
The degraded sinogram $\hat{s}$ is reconstructed to obtain an intermediate image $I$.
The image-domain degradation is defined as in equation (\ref{eq:image_domain}).

\begin{equation}
    \label{eq:image_domain} 
    \hat{I}=\mathcal{G}(I; r_1, r_2),
\end{equation}

\noindent
where $r_1$ and $r_2$ are parameters that determine the degree of CT value adjustment for each region. For $r_1$ and $r_2$, a value of $1.0$ corresponds to no degradation, and smaller values result in stronger degradation.

\begin{figure*}[!t]
    \centering
    \includegraphics[width=\textwidth]{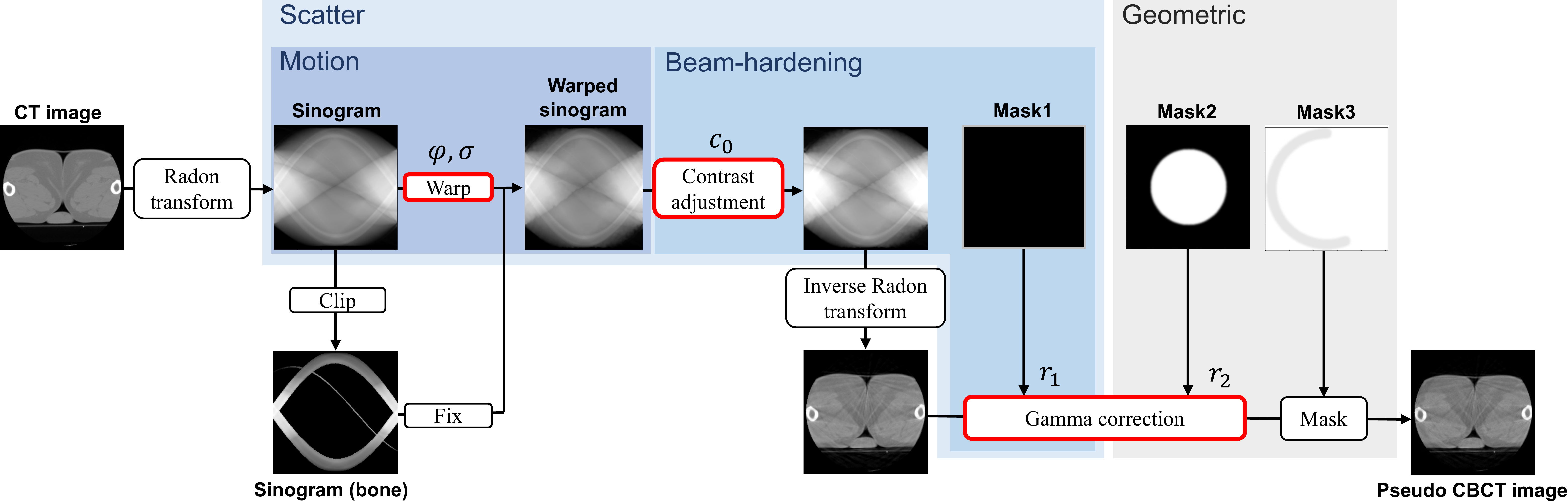}
    \caption{Procedure for creating a pseudo-CBCT image. Background colors indicate the assumed factors of artifacts modeled. The processes shown in red boxes have several parameters that modify the image features. The generated images have the imaging features found in the real CBCT images, while their anatomical structures are consistent with those of the original CT images.
    }
    \label{fig:make_pseudoCBCT}
\end{figure*}

The overall procedure is illustrated in Figure \ref{fig:make_pseudoCBCT}, and the details are described as follows.

\begin{description}
    \item[\textbf{STEP1}] \; Obtain sinogram from CT image by Radon transform
    \item[\textbf{STEP2}] \; Deform the sinogram
    
    \begin{itemize}
        \item [\textbf{1)}] Clip only the bone portion from the sinogram
        \item [\textbf{2)}] Create a displacement field $\varphi$ following a normal distribution $\mathcal{N}(0,\sigma)$ and apply Gaussian smoothing
        \item [\textbf{3)}] Warp the sinogram acquired in STEP1 based on $\varphi$, using VoxelMorph \cite{Balakrishnan2019} 
        \item[\textbf{4)}] Combine the sinogram obtained in 3) with the sinogram of the bone clipped in 1) to obtain a warped sinogram
    \end{itemize}

    \item[\textbf{STEP3}] \; Adjust the warped sinogram contrast with $c_0$
    \item[\textbf{STEP4}] \; Reconstruct the image by inverse Radon transform
    \item[\textbf{STEP5}] \; Adjust the CT values of the reconstructed image
    
    \begin{itemize}
        \item [\textbf{1)}] Gamma correction of CT values in the mask 1 range (entire image) with $r_1$
        \item [\textbf{2)}] Gamma correction of CT values in the mask 2 range (outside the circular region in the center of the image) with $r_2$
        \item [\textbf{3)}] Pixel value shift at the edge of the imaging field with mask 3
    \end{itemize}
    
\end{description}

\noindent
STEP2, STEP3, and STEP5 1) are designed to simulate scatter, beam-hardening, and motion artifacts, whereas STEP5 2) and 3) account for CT value variations due to geometric factors.
For STEP2, the bone portion of the sinogram is combined with the warped sinogram, which prevents deformation of the bone portion of the sinogram and the generation of excessive artifacts around the bone. 
After STEP3, each value of $\hat{s}$ is clipped to the maximum value of $s$.
In STEP5, the pixel values are normalized to [0,1], and the values are returned to the range of the CT values ($-$1000 to 1000 Hounsfield units [HU]) after adjustment.

\begin{figure}[!t]
    \centering
    \includegraphics[width=\columnwidth]{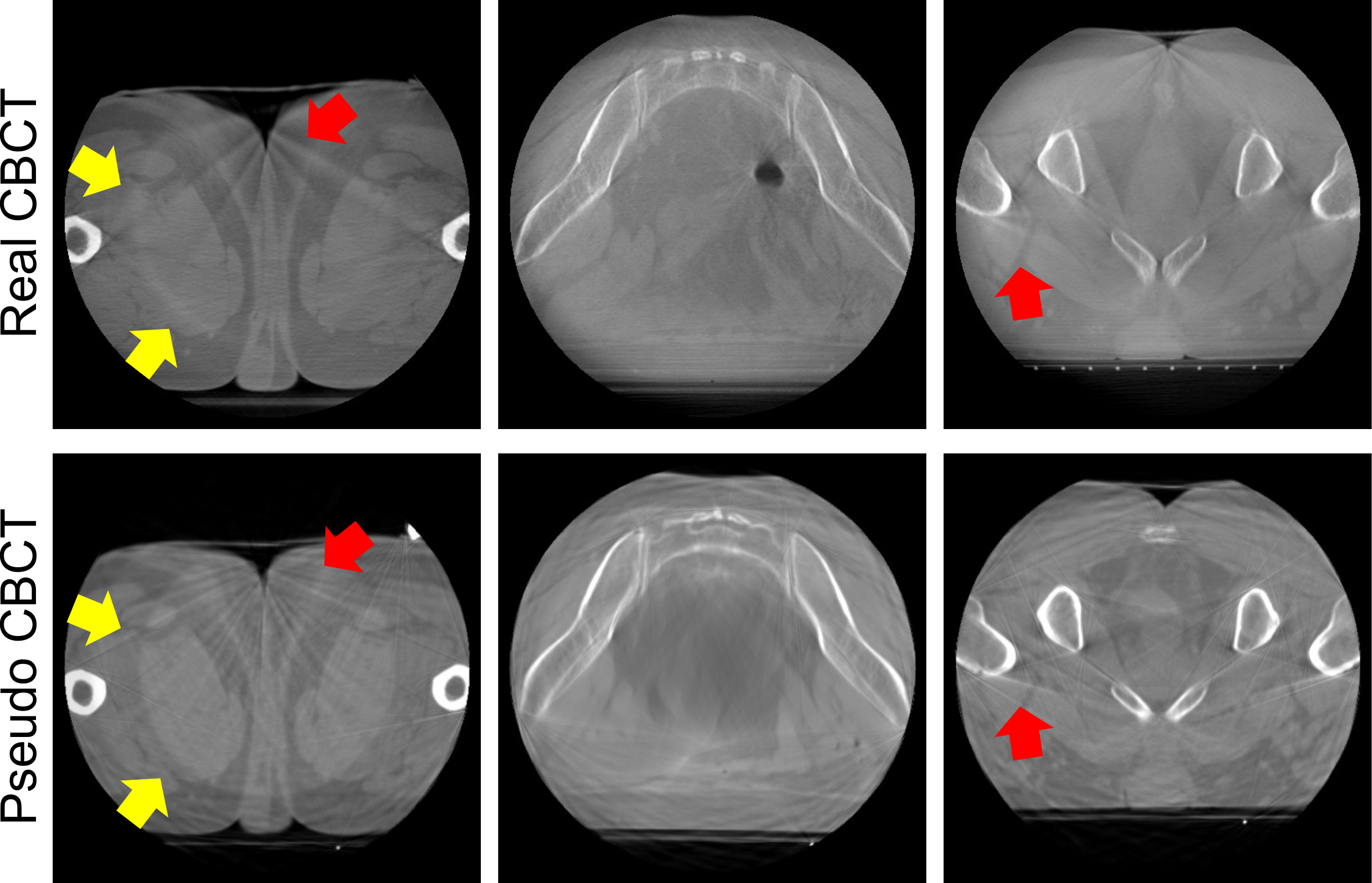}
    \caption{Three examples of real and pseudo-CBCT images. The arrows indicate image characteristics common to both images.
    }
    \label{fig:pseudoCBCT}
\end{figure}

In the processes shown with red boxes in Figure \ref{fig:make_pseudoCBCT}, switching the parameters randomly enables the creation of pseudo-CBCT images with different image characteristics. Examples of pseudo-CBCT images are shown in Figure \ref{fig:pseudoCBCT}. 
The left image is an example without contrast adjustment, in which a decrease in CT values bordering the circular region (yellow arrows) and radial artifacts (red allow) are recreated. On the other hand, the middle image is a contrast-adjusted example that shows contrast defects throughout the image. The right image is also contrast-adjusted, but the $\sigma$ was varied, allowing the artifacts around the bone to be recreated.

\subsection{Conditional latent diffusion model}
\label{subsec:CLDM}
This section describes the conditional latent diffusion model used in the proposed overcorrection-free image translation.
The conditional latent diffusion model used in this study was built by extending the image space-based conditional diffusion model Palette \cite{Saharia2022} on the basis of the structure of the latent diffusion model proposed by Rombach \emph{et al} \cite{Rombach2022}.

The diffusion model \cite{Ho2020} consists of two processes: forward diffusion and reverse denoising. 
In the forward process, noise is gradually added to the image until it becomes nearly pure noise. The reverse process performs the opposite, step-by-step denoising to recover the image. By learning the reverse process, the model can generate images from Gaussian noise. 
In contrast, the conditional diffusion model Palette inputs an image as a condition for each step of the reverse process. This conditioning changes the role of the model from image generation to image-to-image translation, where the output is guided by the structure of the conditional image.
The proposed conditional latent diffusion model (CLDM) attaches an encoder and a decoder to Palette. This allows the diffusion model to be applied over a compressed latent space, thereby reducing computation costs.

\begin{figure*}[!t]
    \centering
    \includegraphics[width=\textwidth]{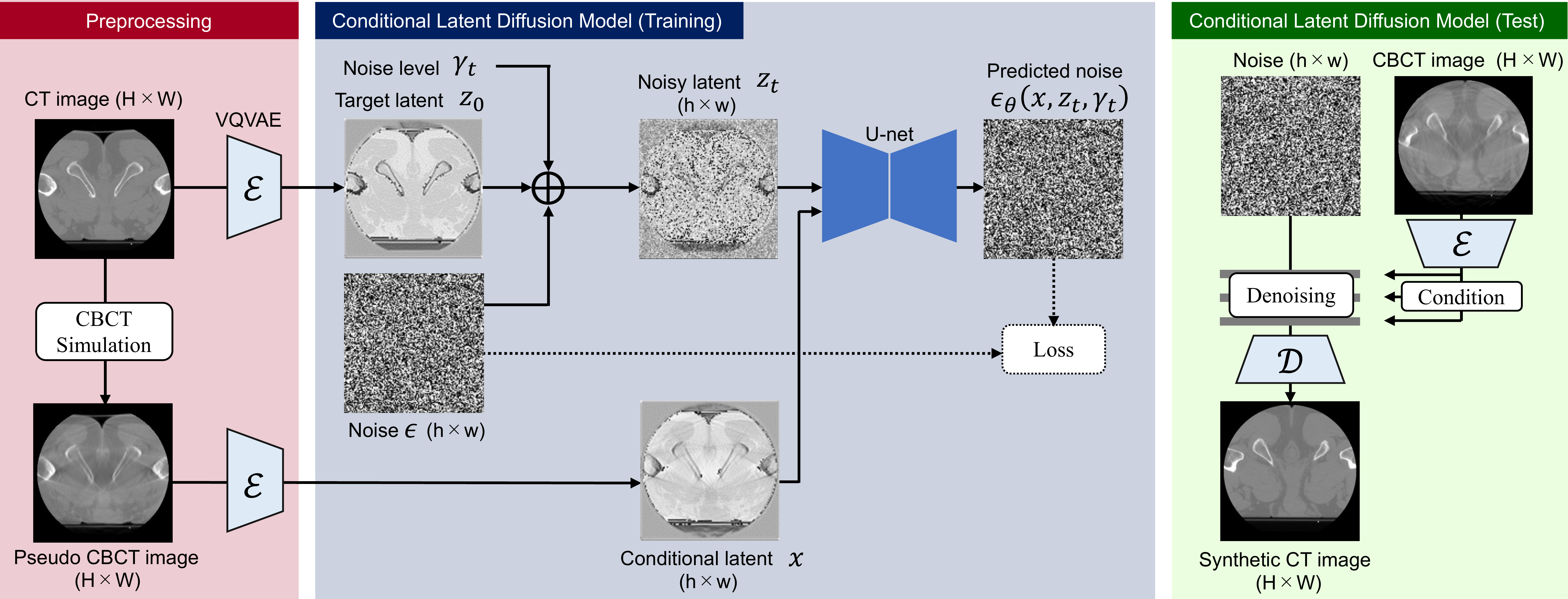}
    \caption{Structure and data flow of the proposed method. Pretrained VQVAE is used for the encoder $\mathcal{E}$ and the decoder $\mathcal{D}$.%
    }
    \label{fig:flamework}
\end{figure*}

The structure and data processing flow of the proposed method are shown in Figure \ref{fig:flamework}. 
Let $\mathcal{E}$ and $\mathcal{D}$ denote the encoder and decoder.
The latent variable obtained from the CT image is denoted as target latent $z_0$, and that from the pseudo-CBCT image as conditional latent $x$.

For encoding and decoding, a pretrained perceptual compression model with fixed gradients is used. Representative perceptual compression models include Variational Auto Encoder (VAE) \cite{Kingma2013}, Vector Quantized VAE (VQVAE) \cite{Oord2017}, and Vector Quantized GAN (VQGAN) \cite{Esser2021}. 
VAE tends to blur output images, and VQGAN incorporates the GAN framework, which may causes learning stability. 
To avoid these issue, we used VQVAE in the proposed method. 
In VQVAE, latent variables acquired by convolution are replaced (vector quantized) with the nearest vector among a finite number of $D$-dimensional vectors, called the codebook. 
The dimension $D$ is a hyperparameter, and was set to 1 in this study.
Thus, the size of the latent variable can be expressed as $h \times w$. $h$ and $w$ are expressed as $h=H/f$ and $w=W/f$, where $H \times W$ is the size of the CT and CBCT images and $f$ is the compression factor.

CLDM learning is performed by adding and predicting noise at randomly sampled timesteps $t (1 \leq t \leq T)$. 
$T$ is a hyperparameter that determines the endpoint of the forward diffusion steps; in this study, as in \cite{Ho2020}, $T=1000$ was used. 
The specific learning method involves Gaussian noise $\epsilon$ being added to the target latent $z_0$ at timestep $t$ and a noisy latent $z_t$ being acquired. The model learns to predict the noise in $z_t$ under condition $x$. The loss function is defined as in equation (\ref{eq:loss}) below, and learning is performed to minimize the squared error between the predicted noise and the added noise $\epsilon$. 

\begin{equation}
    \label{eq:loss} 
    Loss=\mathbb{E}_{x,z_0,\epsilon \sim \mathcal{N}(0,1), t} \| \epsilon_\theta(x, z_t, \gamma_t) - \epsilon\|_2^2,
\end{equation}

\noindent
where $\epsilon_\theta$ is a neural network, and U-Net \cite{Ronneberger2015} is used in the method proposed in Palette. In addition, $z_t$ is defined as in the following equation (\ref{eq:noisy_latent}). 

\begin{equation}
    \label{eq:noisy_latent}    
    z_t=\sqrt{\gamma_t}z_0+\sqrt{1-\gamma_t}\epsilon,
\end{equation}

\noindent
where $\gamma_t$ is a function that determines the noise level at timestep t and is expressed using the noise scheduler $\beta_t$, as in equation (\ref{eq:noise_level}) below. 

\begin{equation}
    \label{eq:noise_level}
    \gamma_t=\prod_{k=1}^t(1-\beta_k).
\end{equation}

\noindent
The two most common noise schedulers $\beta_t$ are the linear scheduler \cite{Ho2020} and the cosine scheduler \cite{Nichol2021}. Among these, the cosine scheduler adds noise gradually and retains information about the target almost until the end of the forward process, enabling efficient learning.
Thus, we used the cosine scheduler in this study, and $\beta_t$ is expressed as in Equation (\ref{eq:scheduler}).

\begin{equation}
    \label{eq:scheduler}
    \begin{aligned}[b]
        &\beta_t=\min\left(1-\frac{\bar{\alpha}_t}{\bar{\alpha}_{t-1}}, \delta\right),\\
        &\bar{\alpha}_t=\frac{f(t)}{f(0)},
        \quad f(t)=\cos\left({\frac{t/T+\tau}{1+\tau}\cdot}\frac{\pi}{2}\right)^2,
    \end{aligned}
\end{equation}

\noindent
where $\delta = 0.999$ and $\tau = 0.008$, following the original setting.

In the testing phase, a latent variable corresponding to a pseudo-CBCT or real CBCT image is input as a condition. 
Under this condition, an image is generated by denoising from Gaussian noise (initial noise) and decoding. 
Since denoising is performed over multiple steps, it takes a reasonable amount of time to generate an image. 
It is possible to reduce the generation time by using accelerated sampling methods \cite{Song2020, Lu2022, Lu2022dpm, Zhao2023UniPC}, but these methods may affect the quality of the generation, and it is not clear to what extent they would affect the improvement in image quality in this study. 
Considering this, images were generated using the same 1000-step setting as in training to evaluate the model's baseline performance and generation time.

\begin{figure}[!t]
    \centering
    \includegraphics[width=\columnwidth]{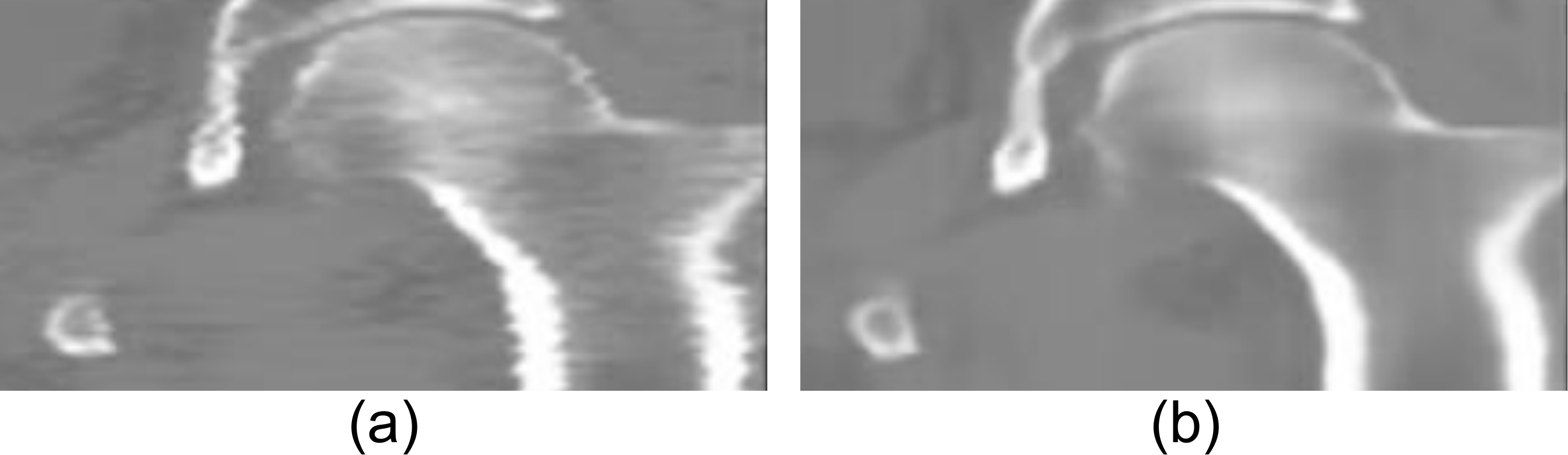}
    \caption{Enlarged coronal views of volumes generated with different initial noise settings. (a) Different initial noise for each slice, (b) The same initial noise for all slices. Using different initial noise introduces inter-slice discontinuities.
    }
    \label{fig:cornal}
\end{figure}

A well-trained model can generate images from arbitrary Gaussian noise. However, the characteristics of the generated image tend to depend on the initial noise \cite{Mao2023}. 
This dependency is visualized in Figure \ref{fig:cornal}, where using different initial noise for each slice causes inter-slice discontinuities, whereas using the same initial noise preserves the three-dimensional continuity. Therefore, images were generated from the same initial noise for all slices in this study.

\section{Experiments}
\label{sec:experiments}
In this study, three experiments were conducted to investigate the performance of the proposed method in improving CBCT image quality.
In Experiment 1, we compared the performance of models with different compression factors in terms of image quality improvement and computation time for pseudo-CBCT images. In Experiment 2, we investigated the model's performance in improving real CBCT images, while also comparing it with conventional methods in CBCT-to-CT image translation. 
In Experiment 3, we performed an ablation study to investigate the effectiveness of each step in the proposed pseudo-CBCT image creation procedure for improving the image quality of real CBCT images.

Python 3.9.7 and Pytorch were used to implement the proposed models. The experiments were performed on a computer with an Intel Core i7-12700F CPU, 64.0 GB of memory, and a 48 GB NVIDIA RTX A6000 GPU. 
The batch size in training was set to 2.
The parameters were set as $\sigma = 8, 16, 24,$ $c_0 = 1.0, 1.15, $ $r_1 = 0.75, 0.85, 0.90, 0.95,$ and $r_2 = 0.85, 0.90, 1.0$. Pseudo-CBCT images were generated by combining these values, with $r_1$ and $r_2$ selected depending on $c_0$.
These values were determined by the experimenter on the basis of visual inspection of the created images and judgment as to whether they recreated the characteristics of the real CBCT images.

\subsection{Dataset}
\label{subsec:dataset}
This study used pelvic CT and CBCT images from 75 patients with prostate cancer who received radiotherapy at Kyoto University Hospital. This study was approved by our institutional review board (approval number: R1446-2). For each patient, the data set included planning CT images and spotlight CBCT images taken on the day of treatment.

Because of differences in the date and time of the imaging and imaging conditions, the following differences may be present between the CT and spotlight CBCT images. 

\begin{itemize}
    \item Imaging range: CT images include the entire body contour, but spotlight CBCT images have a restricted field of view and lack part of the contour.
    \item Resolution
    \item Patient posture
    \item Location and shape of organs
\end{itemize}

\noindent
To reduce these differences, we performed the same preprocessing as used in a previous study \cite{Hase2021}. Specifically, using rigid registration with CT images as the source and CBCT images as the target, the number of slices and resolution of the CT images were adjusted to those of the CBCT images. To focus on improving the image quality, the images were cropped to a 25-cm-diameter circular region centered on the prostate in the transverse section of the body axis to match the imaging range of the spotlight CBCT. From the CT images with these preprocessing steps, pseudo-CBCT images were created using the procedure described in Section \ref{subsec:make pseudo-CBCT}.

Through the above process, CT-real CBCT image pairs and CT-pseudo-CBCT image pairs (both 512$\times$512 pixel, 51–88 axial slices) were obtained for 75 cases. Of these, 66 randomly selected cases (4925 slices) were used as training data, one (59 slices) was used as validation data, and the remaining eight cases (613 slices) were used as test data. 
The pixel values [$-$1000 HU, 1000 HU] were normalized to [0, 1], and the data were entered into VQVAE. Since the input to the diffusion model section needs to be normalized to [$-$1, 1], the latent variables before quantization were normalized by the maximum and minimum values of the pretrained codebook to bring their values close to [$-$1, 1].

\subsection{Preprocess}
\label{subsec:preprocess}
In this section, we describe the training methods of VQVAE and its reconstruction performance. VQVAE was trained with three different compression factors that changed the number of convolution layers. These compression factors were 2, 4, and 8, and the VQVAE for each factor is hereafter denoted by the size of the latent variable after compression. For example, VQVAE with a compression factor of 4 is denoted as VQVAE (128). 
Three types of images (CT, real CBCT, and pseudo-CBCT) were used as data. The maximum number of epochs in training was set to 1000 for each VQVAE, and EarlyStopping \cite{Morgan1989} with a patience parameter of 50 epochs was used.

The reconstruction performance of VQVAE was evaluated numerically using quantitative evaluation indices, and visually through manual observations, applied to the reconstructed images. The quantitative evaluation indices used were the mean absolute error (MAE) and structural similarity (SSIM) \cite{Wang2004} between the original and reconstructed images.

\begin{figure}[t]
    \centering
    \includegraphics[width=\columnwidth]{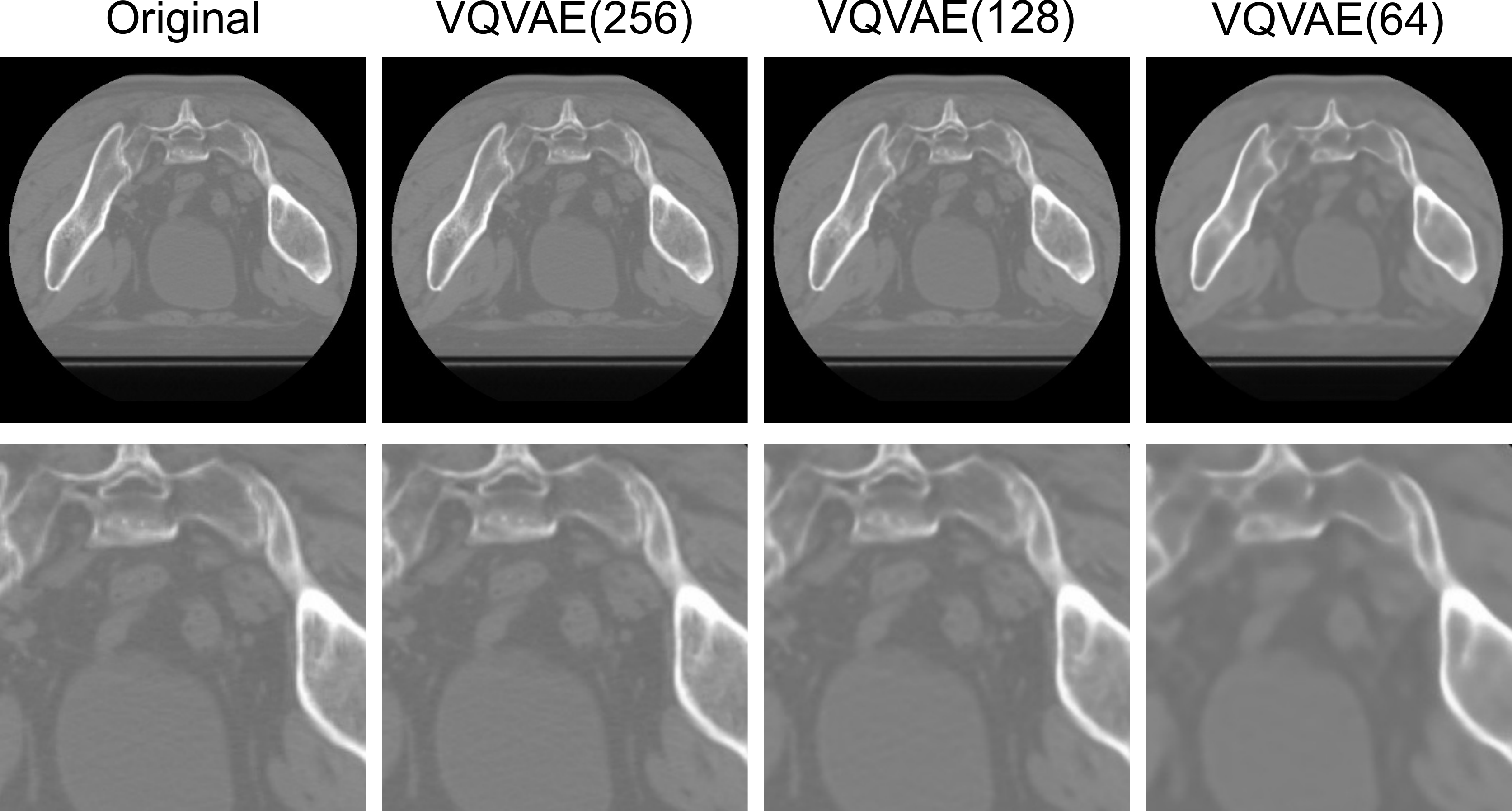}
    \caption{Comparison of reconstruction results of VQVAEs. The upper panel shows the original image and the reconstructed image at each compression factor. The lower panel provides magnified regions for assessment of fine detail.
    }
    \label{fig:prepocess}
\end{figure}

Table \ref{tab:preprocess} shows the average values of MAE and SSIM between the original image and reconstructed image at each compression factor. 
\begin{table}[t]
    \caption{Reconstruction accuracy of VQVAE at each compression factor.
    }
    \label{tab:preprocess}
    \centering
    \begin{tabular}{lcc}
        \hline
        & MAE [HU] & SSIM\\
        \hline
        VQVAE (256) & $2.70$ & $0.998$\\
        VQVAE (128) & $6.08$ & $0.988$\\
        VQVAE (64) & $12.0$ & $0.961$\\
        \hline
    \end{tabular}
\end{table}
VQVAE (256) had the lowest MAE value and the highest SSIM value, and the reconstruction performance decreased with increasing compression factor.
Qualitatively, from an example of the reconstructed images shown in Figure \ref{fig:prepocess}, it can be observed that all VQVAEs reconstructed the original image. 

However, when focusing on enlarged images of fine details, noticeable blurring is observed for VQVAE (64), including unclear tissue contours and smoothing of internal bone structures, making it difficult to regard the image reconstruction at this compression level as sufficient.

\subsection{Evaluations on pseudo-CBCT images and model comparisons}
\label{subsec:ex1}
In Experiment 1, each of the proposed CLDMs with different compression factors were compared in terms of image quality improvement for pseudo-CBCT images and computation time. We compared four models: a conditional diffusion model without compression (Palette) and CLDMs with compression factors of 2, 4, and 8. Hereafter, as in VQVAE, each CLDM will be denoted as CLDM (256), CLDM (128), or CLDM (64), respectively, based on the size of the latent variable. For the compression of each model, the VQVAE trained in Section \ref{subsec:preprocess} was used.

\subsubsection{Experimental and evaluation methods}
For the experimental methods, four models with different compression factors (Palette, CLDM (256), CLDM (128), and CLDM (64)) were trained with CT-pseudo-CBCT image pairs.
Each model was trained for 200 epochs, based on the results of a preliminary experiment to validate the model performance at each epoch, confirming that this number of epochs was sufficient for convergence of the CLDM (128).

In the test phase, the models trained at the 200th epoch were used. 
The initial noise was set on a model-by-model basis, and all slices were generated from the same noise. 
For each model, the initial noise was selected from 100 candidates generated using different seed values, and the one that achieved the best performance on the validation data was used.

For the quantitative evaluation, we calculated MAE and SSIM between the generated and ground truth images. Moreover, the performance of each model was also evaluated in terms of training and generation efficiency by measuring the computation time per epoch for training and per slice for generation.

\subsubsection{Results}
Table \ref{tab:ex1} shows the average MAE and SSIM between the ground truth (reference CT) and original (pseudo-CBCT) image, and between the ground truth and each model's generated image, as well as the computation time per unit for each model. 
\begin{table}[t]
    \caption{Comparison of quantitative evaluation results for image quality improvement of pseudo-CBCT images and computation time.
    }
    \label{tab:ex1}
    \centering
    \begin{tabular}{lcccc}
        \hline
        & & &\multicolumn{2}{c}{Computation time}\\ 
        \cline{4-5}
        & MAE [HU] & SSIM & Training & Generation \\
        & & & [min/epoch] & [sec/slice] \\
        \hline
        Pseudo CBCT & $74.0$ & $0.902$ & & \\
        \bhline{0.03em}
        Palette & $69.0$ & $0.684$ & $52$ & $140$\\
        
        CLDM (256) & $19.5$ & $\bf{0.955}$ & $15$ & $40$\\
        
        CLDM (128) & $\bf{17.4}$ & $0.946$ & $\bf{6}$ & $\bf{30}$\\
        CLDM (64) & $21.8$ & $0.924$ & $\bf{6}$ & $\bf{30}$\\
        \hline
\end{tabular}
\end{table}
CLDM (128) had the lowest MAE value, followed by CLDM (256) and CLDM (64). For SSIM, CLDM (256) had the highest value, with the value decreasing as the compression factor increased. Meanwhile, Palette performed clearly worse than the other models in both MAE and SSIM.
In terms of computation time, CLDM (128) and CLDM (64) had the shortest computation time for both training and generation, at 6 minutes per epoch in training and 30 seconds per slice for generation. In particular, their learning time was less than half that of CLDM (256) and less than 1/8 of Palette, which had the longest computation time.
These results show that models with higher compression ratios can achieve faster training while achieving sufficient improvement, even under constrained training settings and limited computational resources.

\begin{figure*}[t]
    \centering
    \includegraphics[width=\textwidth]{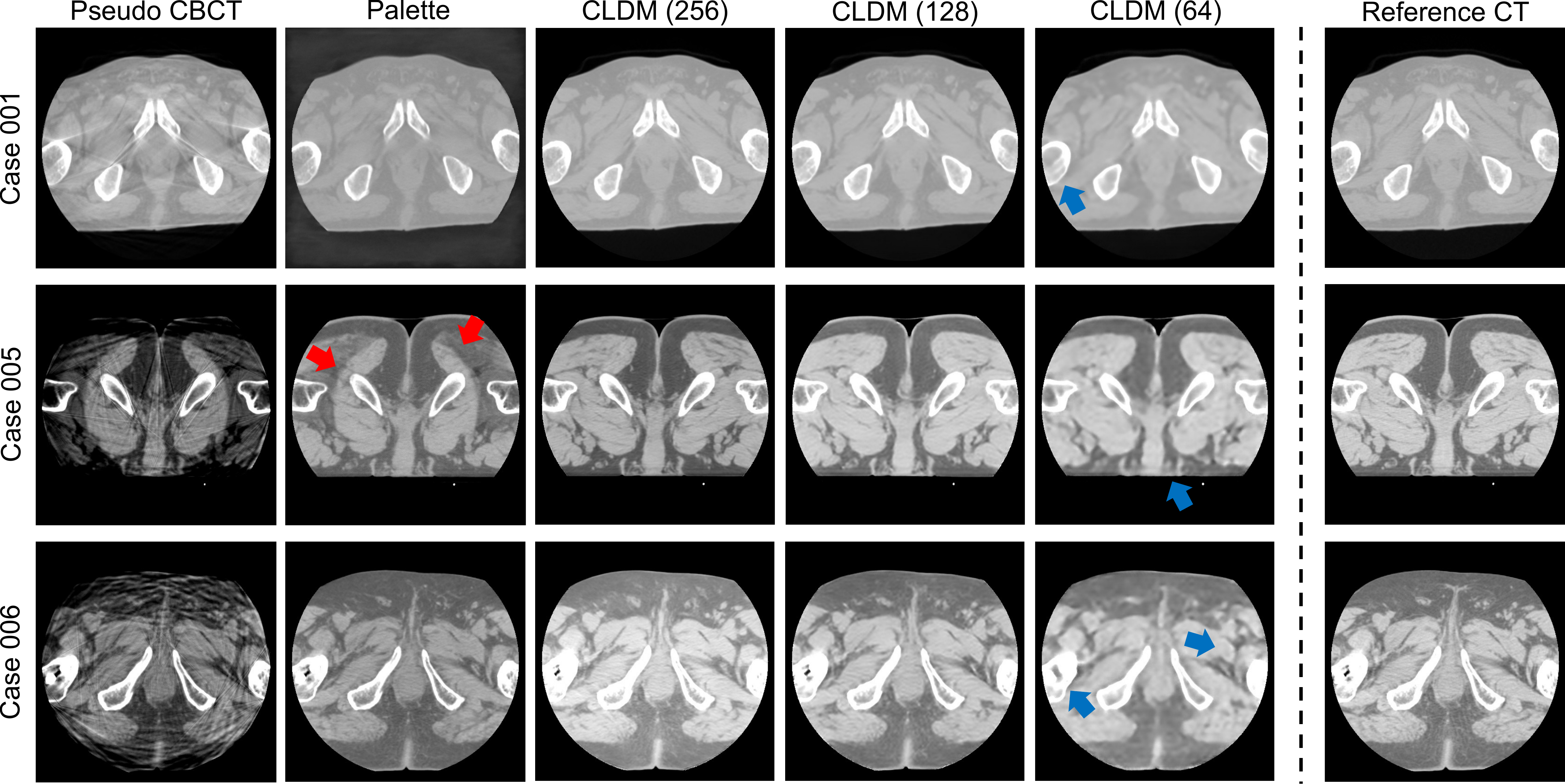}
    \caption{Comparison of image quality improvement for pseudo-CBCT images. The display window range was set to [$-$1000 HU, 1000 HU] for case 001, and [$-$300 HU, 150 HU] for cases 005 and 006. Red arrows indicate residual artifacts, and blue arrows show examples of blurred areas. Except for CLDM (128), there are some errors in CT values and insufficient improvement.
    }
    \label{fig:ex1}
\end{figure*}

Figure \ref{fig:ex1} shows three examples of translation for pseudo-CBCT images for each model. 
In Palette, noise in the air region (case 001), residual circular artifacts (case 005, red arrows), and overall darkening compared with the reference image (case 006) are observed.
In CLDM (256), there are no residual artifacts or noise, but the results are slightly darker (case 005) and slightly brighter (case 006) relative to the reference image. These residual artifacts and the deviations in CT values are probably caused by insufficient learning. No residual artifacts are observable in the images from CLDM (128) and CLDM (64). However, blurring can be seen throughout the CLDM (64) image, including the areas shown by the blue arrows, and likely due to the reconstruction performance of VQVAE. 
This blurring includes tissue contours and internal bone structure, and it is difficult to say that high-quality improvement has been achieved. 
Based on the above experiments, we considered CLDM (128) to be the best model, achieving high-quality image generation with low computational cost.

\subsection{Evaluation on real CBCT images and comparisons with conventional methods}
\label{subsec:ex2}
In Experiment 2, we investigated the performance of the model trained in Experiment 1 for improving the image quality of real CBCT images, and compared its performance with conventional methods. CLDM (128), which performed the best in Experiment 1, was adopted as the proposed CLDM (Pseudo). Three models were used for comparisons: GAN-based (Real) and CLDM (Real), both trained on CT-real CBCT image pairs, and GAN-based (Pseudo), trained on CT-pseudo-CBCT image pairs as in the proposed method. Here, GAN-based refers to the GAN-based image translation model using paired data with cycle consistency loss proposed in a previous study \cite{Hase2021}.

\subsubsection{Experimental and evaluation methods}
\label{subsubsec:ex2-1}
First, we describe the experimental methods for each model. For CLDM (Real), the experimental conditions were the same as for CLDM (Pseudo), except that it was trained on CT-real CBCT image pairs.  GAN-based (Real) and GAN-based (Pseudo) were trained, and the models at 200 epochs—just before the learning became unstable—were used for test. All four models were tested using real CBCT images.

Next, we describe a method for quantitatively evaluating the improvement in image quality for real CBCT images. In this study, we conducted two evaluations, one on the structural change and the other on the CT value distribution, from the viewpoints of maintaining structure and CT value conversion, respectively, which are required to improve the quality of CBCT images.

In the evaluation of structural changes, whether or not the generated images (Synthetic CT images, SynCT images) maintained the anatomical structure in the real CBCT images was evaluated by calculating the error between SynCT and the real CBCT image. The specific calculation procedure is shown below. 

\begin{description}
    \item[\textbf{STEP1}] \; Create a difference image between a SynCT and real CBCT image for each slice.
    \item[\textbf{STEP2}] \; For the difference image, replace the pixel values in the area where the absolute error is less than the threshold value with 0. 
    \item[\textbf{STEP3}] \; Calculate the Root Mean Squared Error (RMSE) in the region within the field of view for the thresholded image acquired in STEP2.
    \item[\textbf{STEP4}] \; Count the number of errors (non-zero pixels, i.e., pixels where overcorrection has occurred) in the thresholded image. 
   
\end{description}

\noindent
The difference image created in STEP1 includes CT value changes caused by image quality improvement. To prevent these changes from affecting the evaluation, we focused on extreme CT value changes associated with structural changes in the bone and air. 
Specifically, thresholding in STEP2 was used to exclude, as much as possible, CT value changes resulting from image quality improvement.
The threshold value was determined after examining several values. The thresholded images were visually checked, and 600 HU, which was considered to have the smallest effect on the CT value change due to improvement, was used as the threshold value.

In the evaluation of the CT value distribution, we assessed how closely the CT values of the improved image approached those of the reference CT image, both for the entire image and for partial regions.
For the entire image, we used the average image to evaluate the CT values because the structural differences between images have a large impact on the evaluation if the evaluation is performed simply on a per volume basis. For each case, average images were created for real CBCT, SynCT, and reference CT images, and the mean and standard deviation of the CT values were calculated. 
In addition, histograms of the CT value distributions in the average images were created, and correlation coefficients between the histograms of reference CT and those of real CBCT and SynCT were calculated within [$-$500, 500 HU], where the histogram peaks existed. 
For partial evaluation, an example-based evaluation was performed. The average CT values within each 4 × 4 pixel region of interest (ROI) were compared among real CBCT, SynCT, and reference CT images. ROIs were placed at identical locations in the real CBCT and SynCT images, and in regions contained similar tissue structures in the reference CT images.

\subsubsection{Results}
\begin{figure*}[t]
    \centering
    \includegraphics[width=\textwidth]{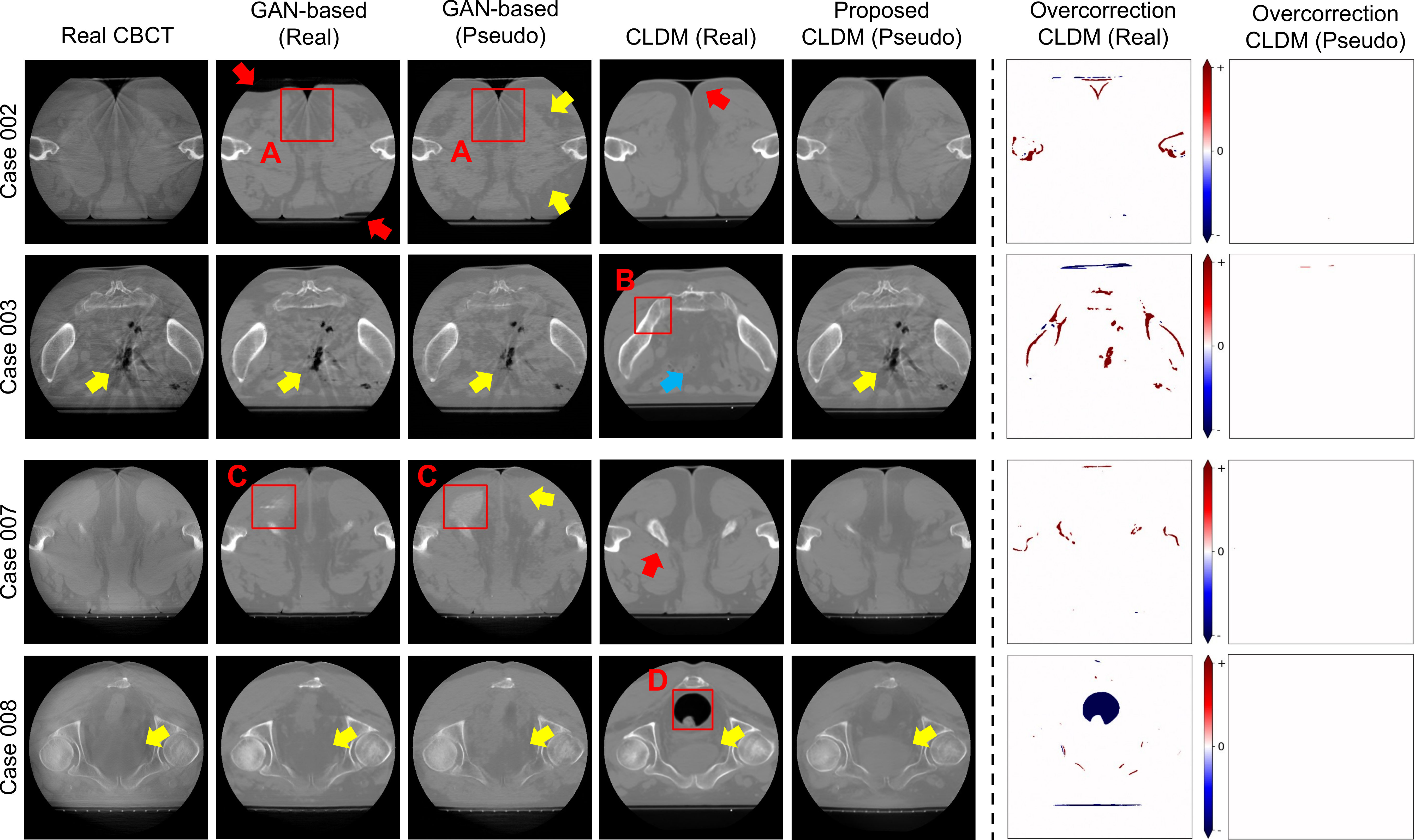}
    \caption{Improvement results for real CBCT images and color maps of overcorrected areas. The annotations and arrows indicate residual artifacts or structural changes. First column: the real CBCT image of each case. Second to fourth columns: images generated by the comparison method. Fifth column: images generated by the proposed method. Sixth and seventh columns: areas of overcorrection with absolute error of 600 HU or more between the real CBCT and the generated image. 
    }
    \label{fig:ex2}
\end{figure*}

Figure \ref{fig:ex2} shows the results of image quality improvement for each method on real CBCT images. 
The figure shows two cases where improvement was confirmed and two cases where improvement was considered insufficient. 
The color maps in the figure are the thresholded images created in Section \ref{subsubsec:ex2-1}, visualizing only the pixels where overcorrection occurred between the SynCT and real CBCT images.
Regions with larger CT values in the SynCT image than in the real CBCT image are shown in red, and regions with smaller CT values are shown in blue. For example, if a region that is soft tissue in the real CBCT image changes to gas in the SynCT image, the region is displayed in blue. 

We first focus on the generated images in Figure \ref{fig:ex2}. The results of both GAN-based methods show residual artifacts, as indicated by annotation A. They also contain unnatural high-intensity regions in the areas indicated by annotation C, which do not exist in the real CBCT image, suggesting the generation of hallucinated structures.
In addition to these observations, the GAN-based (Real) image shows unnatural deformations of the body contour, as indicated by the arrows in case 002, indicating insufficient preservation of shape.
The results of the GAN-based (Pseudo), in contrast, do not show such deformation, but artifacts sometimes remain or are excessively reduced.
Specifically, in case 002, the decrease in CT values bordering the circular region indicated by the arrows is observed. In addition, in case 007, excessive removal of artifacts obscures the tissue contours in the areas indicated by the arrow.

CLDM (Real) effectively reduces artifacts in multiple cases, but this improvement is accompanied by structural changes. 
For example, shape changes in the bony structures are observed in the areas indicated by annotation B and by the arrow in case 007. In addition, the space between the buttock and the fixture, indicated by the red arrow in case 002, is smaller than that in the real CBCT image. Furthermore, topological changes are occasionally observed, with the gas replaced by soft tissue as indicated by the arrow in case 003 and, conversely, the soft tissue replaced by gas as indicated by annotation D.

The proposed method does not exhibit the deformations, topological changes, or hallucinated structures observed in the other methods. However, certain regions remained challenging across all methods.
Specifically, as in the region indicated by the yellow arrow in case 003, areas with complex and clear artifacts around gas made it difficult to improve, and artifact reduction in such regions tended to be only partial. As in the regions indicated by the yellow arrows in case 008, areas with extremely poor contrast, where underlying anatomical structures were obscured, made it difficult to render organ contours. In such regions, although the rendered shapes were different from those visible on the real CBCT images, CLDM-based methods tended to produce clearer organ contours than GAN-based methods.

We then focus on the color maps. 
Only the results of CLDM (Real), which showed many structural changes in the generated images, and those of the proposed CLDM (Pseudo) are shown here as examples.
In the color maps of CLDM (Real), pixels indicated by the red and blue arrows and annotations were extracted from the generated images. In contrast, in the CLDM (Pseudo), the extraction of pixels was almost non-existent. 
These findings are consistent with the qualitative evaluation of the generated images in terms of obvious structural changes related to bone or air.

The quantitative evaluation results of the structural change calculated from the color maps are shown in the left side of Table \ref{tab:ex2}. 
\begin{table*}[t]
    \caption{Quantitative evaluation results for improvement of the image quality of real CBCT images. Average values of eight cases for evaluating structural changes (errors) between real CBCT and synthetic CT (root mean squared error and number of errors in the field of view) and evaluating the distribution of CT values in the entire image (mean $\pm$ standard deviation, correlation coefficient). Examples of average CT values in the region of interest.
    }
    \label{tab:ex2}
    \centering
    \begin{tabular}{lccccccccc}
        \hline
	& RMSE & Number of & Mean $\pm$ SD 
        & Correlation &\multicolumn{5}{c}{ Average within ROI [HU]}\\ 
                \cline{6-10}
                & [HU] 
        & errors [pixels]
        & [HU] 
        & coefficient
        & ROI1
        & ROI2
        & ROI3
        & ROI4
        & ROI5
        \\
        \hline
        Real CBCT & - & - & $-170.2 \pm 262$ & $0.421$ & $-169.5$ & $887.1$ & $177.9$ & $64.1$ & $-97.3$ \\
        \bhline{0.03em}
        GAN-based (Real) & $26.2$ & $598$ & $-87.0 \pm 306$ & $0.908$ & $-58.0$ & $977.0$ & $489.0$ & $26.1$ & $8.5$ \\
        GAN-based (Pseudo) &  $1.21$ & $\bf{3}$ & $-72.5 \pm 272$ & $0.822$ & $-42.1$ & $947.0$ & $255.8$ & $67.9$ & $-109.4$ \\
        CLDM (Real) & $112$ & $3919$ & $-81.9 \pm 315$ & $0.934$ & $-97.1$ & $198.9$ & $69.5$ & $-972.3$ & $-984.7$ \\
        CLDM (Pseudo) & $\bf{1.18}$ & $\bf{3}$ & $-78.0 \pm 283$ & $0.916$ & $-104.0$ & $938.4$ & $73.1$ & $14.8$ & $-128.8$ \\
        \bhline{0.03em}
        Reference CT & - & - & $-77.1 \pm 298$ & $1.00$ & $-97.3$ & $930.4$ & $74.3$ & $35.3$ & $-111.6$ \\
        \hline
    \end{tabular}
\end{table*}
For both RMSE and the number of pixels, GAN-based (Pseudo) and CLDM (Pseudo), which were methods using the pseudo-CBCT images, had very small values, suggesting that the structures were largely preserved. 
In contrast, CLDM (Real) had RMSE about 110 HU higher than that of the proposed CLDM (Pseudo), and the number of pixels was about 1000 times larger, quantitatively investigating substantial structural changes.
These results and the generated images show that the proposed method can enhance image quality while maintaining the structure of the real CBCT image, which is a problem with conventional methods.

\begin{figure}[t]
    \centering
    \includegraphics[width=\columnwidth]{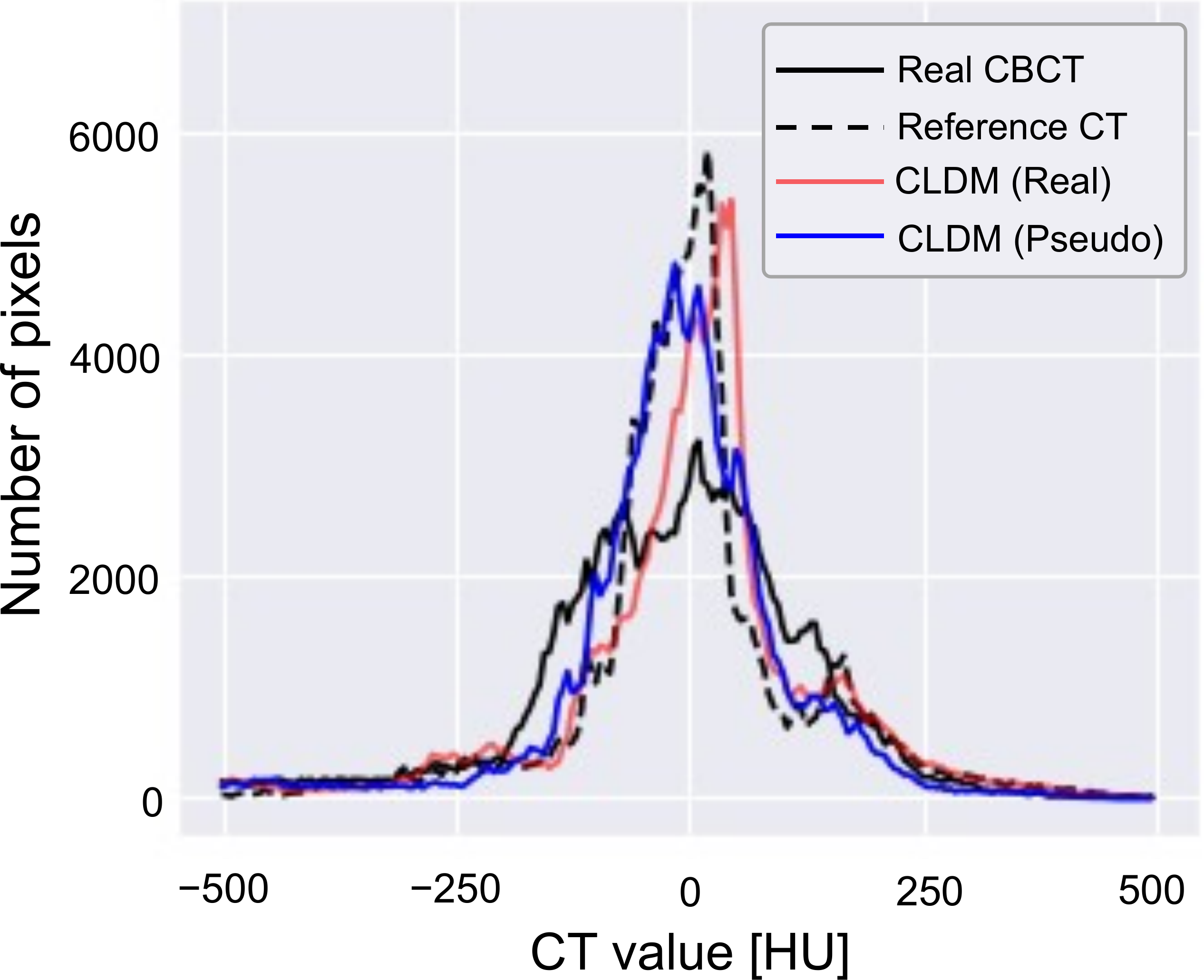}
    \caption{Example histogram distributions of CT values for real CBCT, CLDM (Pseudo), CLDM (Real), and reference CT.
    }
    \label{fig:hist}
\end{figure}

The results of the quantitative evaluations of the CT value distributions for the entire image and for partial regions are shown in the center and right of Table \ref{tab:ex2}, respectively.
For the entire image, the proposed method approached the reference values in terms of the mean and standard deviation of CT values, as well as the correlation coefficients of histgrams. Furthermore, the evaluation values of the proposed method were comparable to those of the GAN-based (Real) and CLDM (Real), which means that the improvement in the CT values achieved by the proposed method is nearly equivalent to that of these conventional methods.
For reference, an example of the histograms is shown in Figure \ref{fig:hist}, indicating that the distributions of the CLDM-based methods shift from that of the original CBCT toward that of the reference CT.
For partial regions, example results for five ROIs (ROI1-ROI5) set within the annotation areas of Figure \ref{fig:ex2} are shown. For each ROI, the proposed method produced values close to the reference. In contrast, GAN-based methods and CLDM (Real) occasionally produced values that deviate considerably from the reference due to overcorrection.

\subsection{Ablation Study}
\label{subsec:ablation study}
In Experiment 3, we performed an ablation study, investigating the effectiveness that each step of the proposed pseudo-CBCT image creation procedure has in improving the image quality of real CBCT images.

\subsubsection{Experimental and evaluation methods}
In this experiment, we used images in which certain steps were omitted from the creation procedure shown in Figure \ref{fig:make_pseudoCBCT}. The following five processes were omitted. 

\begin{itemize}
\item w/o warp
\item w/o contrast (adjustment)
\item w/o mask1
\item w/o mask2
\item w/o mask3
\end{itemize}

\noindent
Each model was trained using the pseudo-CBCT images created by the above procedure and is hereafter denoted by the name of the omitted process. The performance of each model in improving the quality of real CBCT images was compared with that of the proposed method. The experimental conditions except for the training data were the same as those used in the proposed method. For quantitative evaluation, the mean and standard deviation of the CT values and the correlation coefficients were calculated from among the evaluation metrics described in Section \ref{subsubsec:ex2-1}.

\subsubsection{Results}
\begin{figure*}[t]
    \centering
    \includegraphics[width=\textwidth]{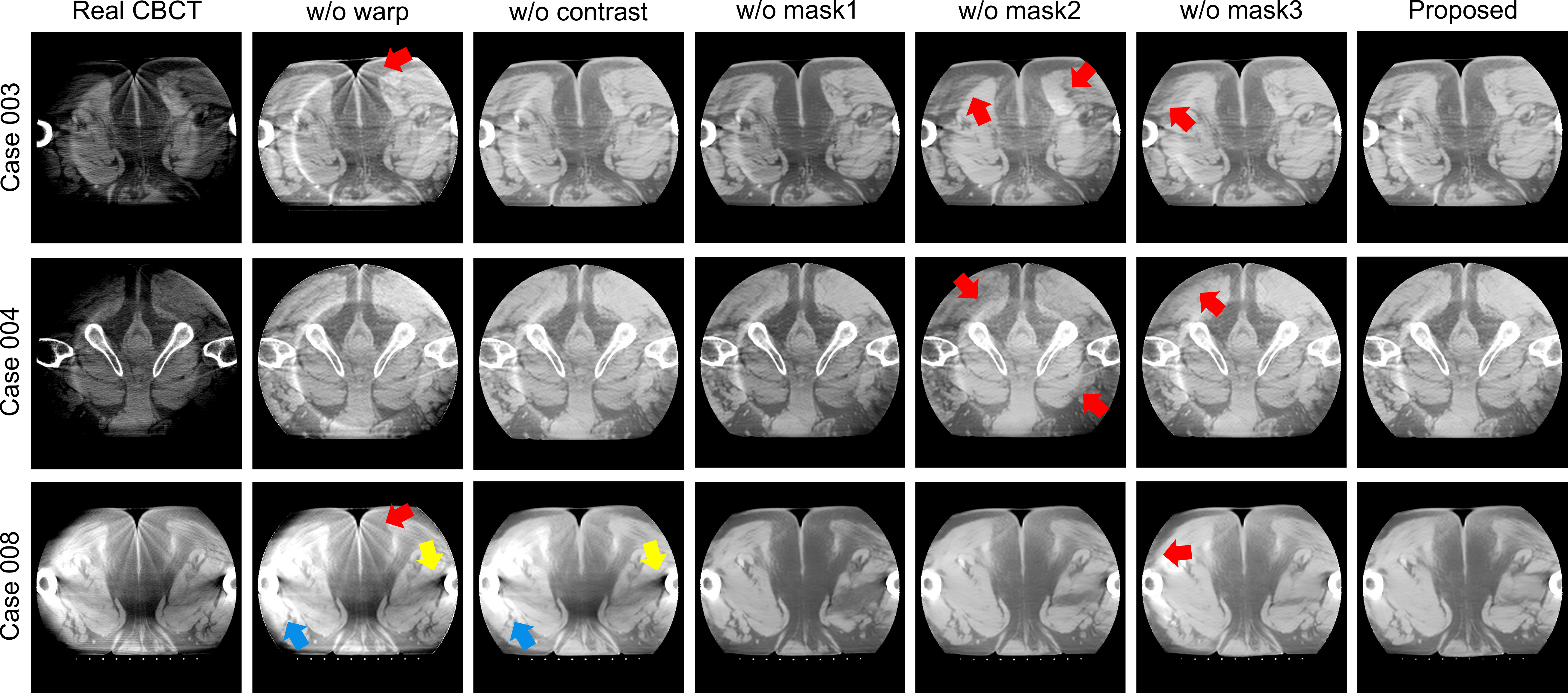}
    \caption{Comparisons of image quality improvement of real CBCT images when changing the pseudo-CBCT image creation procedure. The display window range was set to [$-$300 HU, 150 HU]. In the images obtained by methods other than the proposed method, the CT values were reduced and artifacts remained in the areas indicated by the arrows.
    }
    \label{fig:ablation}
\end{figure*}

The image quality of the real CBCT images was improved for each model trained using the pseudo-CBCT images created by different procedures, with the extent of improvement varying among the models.
The results are shown in Figure \ref{fig:ablation}. Here, the results of the proposed method are used as a baseline for comparisons with the results of the other models.

In the w/o warp model, radial artifacts clearly remain in the areas indicated by red arrows in cases 003 and 008. In case 008, shadow-like artifacts (yellow arrow) and high-intensity artifacts (blue arrow) are also observed around the bone. Similarly, in the w/o contrast model, in which the contrast adjustment of the sinogram is omitted, the artifacts around the bone in case 008 can also be observed, although the radial artifacts have improved.
These results indicate that incorporating operations that consider factors such as scatter, motion artifacts, and beam hardening—even simple ones such as sinogram deformation and contrast adjustment—can contribute to improving the contrast defects and artifacts that interfere with the rendering of anatomical structures in the image.
For the w/o mask1 model, images appear darker than the other models, suggesting that mask1 contributes to correcting the CT value deviations.
Finally, in the w/o mask2 and w/o mask3 models, the CT values in the areas indicated by red arrows are reduced.
These results suggest that these two masks are effective for reducing artifacts associated with decreased CT values caused by geometric factors, which cannot be fully compensated for by sinogram deformation or contrast adjustment alone.

Table \ref{tab:ablation} shows the average values of the quantitative evaluation results. 
\begin{table}[t]
    \caption{Results of the evaluation of CT value distributions (mean $\pm$ standard deviation, correlation coefficient) for different procedures of the pseudo-CBCT image creation process.
    }
    \label{tab:ablation}
    \centering
    \begin{tabular}{lcc}
        \hline
        & Mean $\pm$ SD [HU] & 
        \begin{tabular}{c}
             Correlation \\ coefficient
        \end{tabular}\\
        \hline 
        Real CBCT & $-170.2 \pm 262$ & $0.421$\\
        \bhline{0.03em}
        w/o warp & $-69.6 \pm 278$ & $0.839$\\
        w/o contrast & $-67.8 \pm 288$ & $0.862$\\
        w/o mask1 & $-119.5 \pm 274$ & $0.682$\\
        w/o mask2 & $-91.5 \pm 283$ & $0.903$\\
        w/o mask3 & $-77.9 \pm 285$ & $0.902$\\
        Proposed & $-78.0 \pm 283$ & $0.916$\\
        \bhline{0.03em}
        Reference CT & $-77.1 \pm 298$ & $1.00$\\
        \hline
    \end{tabular}
\end{table}
Both the CT value averages and the correlation coefficients of the proposed method are close to those of the reference image, confirming the effectiveness of the proposed method. 
The correlation coefficients are relatively low for the w/o warp and w/o contrast models, in which artifact persistence was considerable. The w/o mask1 model, which showed a decrease in the overall CT values, showed a large deviation from the reference. 
In contrast, the w/o mask2 and w/o mask3 models, which showed only a partial decrease in CT values, showed a relatively high correlation, but it was not as high as the proposed method. These results quantitatively confirm that all steps of the proposed pseudo-CBCT image creation procedure are necessary to improve the image quality of real CBCT images.

\section{Discussion}
In this study, pseudo-CBCT images with visually reproduced CBCT-like artifacts were generated by replacing CBCT imaging related physical phenomena with simplified processes. Some processes were designed based on geometric factors specific to the spotlight CBCT imaging system, and thus may depend on the dataset. Despite a self-supervised learning with visually simulated pseudo data, the proposed method achieved CT value improvement comparable to that of conventional methods trained on real data, while reducing structural changes in the images. These results suggest that visually reproduced artifacts can be effectively used for training. Since the current pseudo-CBCT creation procedure is based on simplified modeling, further improvement is expected by incorporating additional imaging physics considerations. Such refinement may enhance CT values without overcorrection and improve complex artifacts that remain challenging for both the proposed and conventional methods.

In this study, the field of view of the CT images was limited to a circular region centered on the prostate, matching that of the the spotlight CBCT images used as the target, and image quality improvement was evaluated within this region.
This setting was adopted to ensure consistency with the target data, while the proposed framework itself is not inherently constrained by the field of view.
However, in clinical practice, structural information outside the field of view is essential for accurate treatment. 
We therefore plan to improve on this point by working on both image quality improvement of CBCT images and extension of the field of view in future research.

Because of the memory limitation of the GPU, all experiments were performed on 2D slices. 
In the proposed method, 3D continuity can be preserved even with 2D slice-based processing. However, learning inter-slice relationships may further improve image quality, making extension to 3D processing a promising direction for future work.

\section{Conclusion}
In this study, we proposed a method for enhancing the quality of CBCT images without overcorrection, based on a conditional latent diffusion model using pseudo-CBCT images. 
To overcome the problem of conventional methods, in which anatomical structures differ between input and output images, we performed self-supervised learning using pseudo-CBCT images with simulated CBCT artifacts created by processing CT images, enabling both maintenance of anatomical structure and improvement of image quality.
Furthermore, the framework of the conditional diffusion model was extended to a conditional latent diffusion model to reduce computation costs and improve efficiency.

The results of the evaluations indicate that the proposed method achieved CT value improvement comparable to that of conventional methods, while effectively preserving anatomical structures. The number of pixels affected by structural changes was reduced to less than 1/1000th of that of the conventional method, demonstrating overcorrection-free image translation.
Furthermore, by extending the framework, it was confirmed that the proposed method achieved faster processing while simultaneously performed superior image quality improvement, even under constrained training settings.

\section*{Acknowledgment}
We thank Edanz (https://jp.edanz.com/ac) for editing a draft of this manuscript.

\bibliographystyle{IEEEtran}
\bibliography{reference}



\begin{IEEEbiographynophoto}{Naruki Murahashi} received the bachelor’s degree in human health sciences from Kyoto University, Kyoto, Japan, in 2024, and the master’s degree in human health sciences in 2026. She is currently pursuing the Ph.D. degree with the Department of Human Health Sciences. Her research interests include biomedical engineering and machine intelligence in clinical medicine.
\end{IEEEbiographynophoto}

\begin{IEEEbiographynophoto}{Mitsuhiro Nakamura} received the Ph.D. degree in medicine from Kyoto University, Kyoto, Japan, in 2010. He is currently a Professor with the Medical Physics Laboratory, Graduate School of Medicine, Kyoto University. His research interests include medical physics and high precision radiotherapy. His awards include the Young Scientist Award in Medical Physics of the International Union of Pure and Applied Physics (IUPAP), International Organization for Medical Physics, in 2019.
\end{IEEEbiographynophoto}


\begin{IEEEbiographynophoto}{Megumi Nakao} received the Ph.D. degree in informatics from Kyoto University, Kyoto, Japan, in 2003. He was with the Graduate School of Medicine, Kyoto University, and the Graduate School of Information Science, Nara Institute of Science and Technology, Nara, Japan. He is currently a Professor with the Biomedical Engineering and Intelligence Laboratory, Graduate School of Medicine, Kyoto University. His research interests include biomedical engineering and machine intelligence in clinical medicine.
\end{IEEEbiographynophoto}

\EOD

\end{document}